# Influence of severe plastic deformation on the precipitation hardening of a FeSiTi steel

A. Fillon[1], X. Sauvage[1], A. Pougis[2,4], O. Bouaziz[3], D. Barbier[3], R. Arruffat[2,4], L.S. Toth[2,4]

1- Université de Rouen - Groupe de Physique des Matériaux, UMR CNRS 6634 - Avenue de l'Université, 76801 Saint Etienne du Rouvray Cedex, France

2- Université de Lorraine - Laboratoire d'Etude des Microstructures et de Mécanique des Matériaux, UMR CNRS 7239 - Ile du Saulcy, 57045 Metz, France

3- Arcelormittal Research SA – voie Romaine, 57280 Maizières-lès-Metz, France

4- Université de Lorraine - Laboratory of Excellence on Design of Alloy Metals for low-mAss Structures ('DAMAS') - France

**Corresponding author:**

Xavier Sauvage (e-mail: xavier.sauvage@univ-rouen.fr, tel: +33. (0)2.95.51.42, fax: +33. (0)2.32.95.50.32)

**Abstract** The combined strengthening effects of grain refinement and high precipitated volume fraction (~6at.%) on the mechanical properties of FeSiTi alloy subjected to SPD processing prior to aging treatment were investigated by atom probe tomography and scanning transmission electron microscopy. It was shown that the refinement of the microstructure affects the precipitation kinetics and the spatial distribution of the secondary hardening intermetallic phase, which was observed to nucleate heterogeneously on dislocations and sub-grain boundaries. It was revealed that alloys successively subjected to these two strengthening mechanisms exhibit a lower increase in mechanical strength than a simple estimation based on the summation of the two individual strengthening mechanisms.

**Introduction**

More than 50 years ago, the relationship between the grain size and the yield stress of materials was established by Hall and Petch [1, 2]. The smaller the grain size, the higher the yield stress, thus Ultrafine Grain (UFG) alloys look promising to achieve record mechanical strengths. Although the exact physical mechanisms of plastic deformation in nanoscaled materials are still under debate [3, 4], processing UFG alloys using Severe Plastic Deformation (SPD) [5] has became a fast growing field of research during the past twenty years. Beside the development of viable industrial processes, the usual lack of ductility of SPD materials is another issue that has focused lots of interest [6, 7]. It has been proposed by some authors that a fine distribution of nanoscaled precipitates within the UFG structure could potentially enhance the ductility [8, 9]. Such precipitates may indeed pin dislocations upon deformation, leading to an increase of the dislocation density. Such a scenario would give rise to an increase of the strain hardening and subsequently to an increased ductility. The other advantage of a fine distribution of nanoscaled precipitates would be also a higher yield stress, combining strengthening by grain size reduction effect and the traditional precipitate hardening. It is also interesting to note, that usually second phase particles or precipitates do improve the thermal stability by pinning grain boundaries. This approach was quite successfully pursued by Kim and co-authors in a 6061AA alloy processed by Accumulative Roll Bonding (ARB) [10, 11]. They have shown that a significant strengthening could be obtained thanks to the combination of an ultrafine grained structure and nanoscaled precipitates. Similar features together with an improved ductility were also reported by Zhao and co-authors on a 7075 AA and by Ohashi and co-authors on an Al-11wt%Ag alloy processed by ECAP followed by a precipitation treatment [12-15]. However, controlling the precipitation kinetics in SPD metals is a critical issue because there might be competition between recovery, recrystallisation and grain growth, while heterogeneous precipitation along dislocations and/or grain boundaries is very likely to occur [16]. A very fast rate of precipitation is often reported in aluminium alloys such that all classical metastable phases do not appear [17-20]. In a model FeAuPd alloy, it was also observed that the decomposition process is dramatically affected by the nanoscaled structure resulting from HPT processing [21]. The decomposition process was very different because of the fast grain boundary diffusion that significantly affects the redistribution of the elements.

In this study, we propose to investigate the precipitation hardening of a FeSiTi steel that was severely deformed to achieve a UFG structure prior to the precipitation treatment. FeSiTi steels do exhibit a remarkable hardening potential even with only few percent of Si and Ti [22-24]. The typical aging temperature for the precipitation treatment of FeSiTi alloys is in a range of 400 to 600°C, leading to the nucleation of a high density of spherical and nanoscaled Fe$_2$SiTi particles. These intermetallic particles exhibit a metastable phase (Heusler phase) [25-26] and they are coherent with the bcc Fe matrix. The aim of this work was to investigate the hardening ability and the precipitation kinetic in such steel after grain refinement by Equal Channel Angular Pressing processing (ECAP). A systematic comparison with the conventionally rolled material was also performed.

**Experimental**

Fe-Si-Ti alloy ingots were cast at ArcelorMittal Research S.A. laboratory. The composition of the prepared material as revealed by chemical analysis was as follows (wt%): 2.57Si-1.17Ti-0.007C-0.005Mn-0.003P-0.001B-0.0008S-0.0004N. Ingots were hot rolled between 1230 and 980°C down to a thickness of 3 mm and subsequently water quenched. After this process, the mean grain size was about 25 µm. In the following, this state, corresponding to the solution treated material, will be referenced as the "as-received material". The as-received material was processed by ECAP at room temperature with a 90° die and a back pressure of 125 MPa. Since the material was delivered as 3 mm thick sheets, it was sandwiched between two copper sheets to form a billet of 20 x 20 x120 mm in size. This sandwich was ECAPed up to three passes via route A, corresponding to an equivalent strain of about 3.4. For comparison, the as-received material was also processed by cold rolling (i.e. at room temperature) up to 70% thickness reduction in 10 passes (equivalent strain ε = 1.2). On the basis of previous work on similar alloys, a temperature of 550°C was selected for the precipitation treatment. As-received (reference) and as-deformed (ECAP and rolled) samples were aged in a muffle furnace up to 15 hours.

Vickers microhardness measurements were performed with a 300g load. Before measurements, samples were mechanically polished. Hardness data reported in the present manuscript were obtained from an average of 10 indentations performed at different sample locations.

In FeSiTi alloys, as mentioned above, precipitates are coherent with the bcc Fe matrix and due to misfits they create large lattice distortions making them difficult to image in a conventional TEM. Therefore, microstructures were characterized by Scanning Transmission Electron Microscopy (STEM) and Atom Probe Tomography (APT). STEM specimens were first extracted with a diamond saw, and then mechanically polished. They were electropolished at 16°C in a 90% ethylene glycol monobutyl ether + 10% perchloric acid solution to achieve electron transparency. Samples were finally cleaned by ion milling using a GATAN® PIPS 691 (3 keV, ±3° angle) to remove the thin oxide layer resulting from the electropolishing process. Observations were performed with a JEOL ARM200F microscope operated in the STEM mode with a probe size of 0.2 nm. High

Angle Annular Dark Field (HAADF) images were recorded with collection angles ion in a range of 50 to 180 mrad. Since precipitates contain a large amount of light elements (Si and Ti), they are darkly imaged on HAADF images. APT samples were prepared by standard electropolishing techniques. Analyses were performed in UHV conditions, using an energy compensated atom probe equipped with an ADLD detector [27]. Samples were field evaporated using electric pulses (30kHz pulse repetition rate and 20% pulse fraction) at a temperature of 80K. Such conditions are known to provide quantitative measurements in FeSiTi alloys [28].

**Results**

The evolution of the hardness of the material as a function of the aging time at 550°C is plotted on Fig.1. As expected, due to the nucleation of Si and Ti rich precipitates, a significant increase in hardness was observed for all three different states. For the as-received material (i.e. without deformation prior the aging treatment), an increase in hardness of about 300 MPa was achieved in about 5 hours. The rolled and ECAP samples displayed an initial hardness higher than the as-received material (about 100 MPa higher) due to strain hardening and grain refinement. However, after aging, their peak hardness was significantly below the non-deformed alloy (450 vs 500 MPa). It is interesting to note that these two deformed states exhibited a very similar hardening behaviour upon aging. Beside, their peak hardness was achieved in only 2h instead of 5h (for the non-deformed material) and the increase of hardness was only about 120 MPa instead of 300 MPa (for the non-deformed material). Thus, the resulting strengthening effect appears less marked in the deformed state whatever the processing route and it seems that there was no simple cumulative effect of strain hardening, grain boundary strengthening and precipitate hardening, probably because these contributions were partly masked and counterbalanced by recovery and/or recrystallization phenomena.

The hardness of both the ECAP and the rolled materials decreased by about 10% after 15h at 550°C while no significant change in the non-deformed material was reported. Microstructure characterisation was carried out to clarify these differences. APT analyses were mainly used to determine the mean composition of precipitates and of the matrix while STEM observations provide some information about the precipitate distribution and a better statistics about the mean precipitate size.

APT analyses of the rolled/ECAP processed samples were carried out before aging and after 2h and 6h at 550°C. Before aging (data not shown here), Ti and Si atoms were homogeneously distributed within the matrix and the measured composition was very close to the nominal composition given by chemical analysis (6.01 ±0.04 at.% Si and 1.10 ±0.02 at.% Ti). Impurities like Al, B, P, Cr, V and C were also detected, with a total amount of about 0.3 at%.

Nanoscaled Si and Ti rich precipitates nucleated and grew during aging (Fig. 2). As expected and as shown on the concentration profile computed across two neighbouring precipitates (Fig. 3), the precipitate composition was very close to 25 at% Ti, 25 at% Si and 50 at% Fe, corresponding to the $Fe_2SiTi$ intermetallic phase. The typical precipitate diameter was about 7 nm. It is also interesting to note that some segregation of B, P and Ti along some crystallographic defects was also observed (Fig. 4). Unfortunately, the depth of this analysed volume was quite limited due to an early specimen failure during field evaporation and it was impossible to determine if it had been segregation along a linear defect (dislocation) or a planar defect (grain or sub-grain boundary). Matrix and precipitate mean compositions were determined by data filtering and they are reported in Tab. 1 for all investigated states. One should note that there is no significant difference in the composition of the precipitates in the rolled versus ECAPed alloy. The precipitates have the equilibrium composition of the $Fe_2SiTi$ intermetallic phase after two hours of aging in both cases. The volume fraction $f_v$ of precipitates was estimated using the following equation: $f_v = ( C_0 - C_m ) / ( C_p - C_m )$, where $C_0$ is the nominal concentration of Ti (respectively Si), $C_m$ the Ti (respectively Si) mean concentration in the matrix and $C_p$ the Ti (respectively Si) mean concentration in precipitates. Values given in Tab. 1 are an average of the volume fraction estimated from Ti and Si concentrations. Once more, it clearly appears that differences between the rolled and the ECAPed material were not significant.

In order to get a better statistic on the precipitate size and distribution within the matrix, some STEM observations were carried out. $Fe_2SiTi$ precipitates are darkly imaged on HAADF images because the average atomic number in this phase is lower than in the matrix. As shown in Fig. 5, most of precipitates are spherical shaped but they are not homogeneously distributed, both in the rolled and in the ECAPed material. Obviously some heterogeneous precipitation occurred along crystallographic defects like dislocations, grain and sub-grain boundaries. The mean radius of $Fe_2SiTi$ particles was measured on these HAADF images. It was 4.8 nm and 3.8 nm after 2h aging at 550°C, respectively for the rolling and ECAP states, respectively.

**Discussion**

Results obtained in the present study have confirmed that mechanical properties in the Fe-2.57wt%Si-1.17Ti system could be significantly improved due to strain hardening obtained either by SPD processing ($\Delta HV_{refinement}$~ +100 MPa, obtained after ECAP processing at $\varepsilon = 3.4$ and also after rolling at $\varepsilon = 1.2$) or by precipitation treatment ($\Delta HV_{precipitation}$~ +300 MPa, obtained after 5h at 550°C). Nevertheless, it has been found that when the two preceding strengthening mechanisms were successively combined in Fe-Si-Ti alloys, which included grain refinement by initial SPD processing followed by precipitation hardening treatment, the combining effect did not increase more the hardening potential of the system ($\Delta HV$~ +250 MPa, obtained after SPD processing and 2h aging). The maximal microhardness measured by combining these two strengthening mechanisms remained lower than the maximum precipitation hardening obtained in the non-deformed alloy, leading to a reduced beneficial effect on the mechanical properties of the deformed material, in comparison with the mechanical behaviour obtained from the individual contribution of the precipitation hardening in the non-deformed alloy. In fact, some recovery or recrystallization might have appeared during the heat treatment of the deformed structure, leading to the softening of the as-deformed material. In most cases the hardness increase was due to the combination of several contributions, such as solid solution, grain refinement, precipitation hardening. However, these phenomena are complex, interdependent, and generally not rigorously additive, especially when they take place within the same volume of the material. Therefore, it remains difficult to separate the contribution of each factor to alloy strength. It should be noticed that synergic effects of the combined grain size refinement and precipitation strengthening on mechanical properties of a CuCrZr alloy were discussed in a recent paper [29], and it was shown that strengthening by grain boundaries and precipitates are coupled at the processing stage and subsequently are not a simple and direct sum of the two strengthening mechanisms acting alone. Another important point is that the microstructure produced by SPD processes was observed to greatly accelerate the precipitation kinetics in the Fe-Si-Ti system. Defects induced during the plastic deformation, included dislocation cells, sub-grain boundaries, higher angle grain boundaries, constitute a path for high speed diffusion. The atomic disorder associated to a grain boundary contributes to reduce the activation energy barrier for nucleation and acts as precipitate-forming sites, favouring heterogeneous nucleation of precipitates and accelerating precipitation kinetics. In the case of coherent precipitation, it was found that dislocations also reduce the contribution of the lattice strain energy and, subsequently, decrease more the activation energy barrier for nucleation and also contribute to develop heterogeneity of the precipitates nucleation in the material and accelerated precipitation kinetics [30-32]. It should be expected that the strengthening effect due to precipitation hardening in the as-deformed alloys is very different from the one induced by the homogeneous distribution of the precipitates in bulk (interior) of grain observed in the non-deformed material. Indeed in the present study, the microstructural features observed in the as-deformed alloys have revealed localized precipitation at grain boundaries and on dislocations leading to a lower density of hardened precipitates in the bulk of grain and subsequently, strongly reduced the strengthening contribution of the precipitation hardening in the bulk of grain. This finding provides evidences that the plastic strain in the deformed alloys took place predominantly at grain boundaries, and partly explains the relatively lower strain hardening observed in the as-deformed alloy (compared with the one obtained in the non-deformed material). It is well established that SPD processed materials develop a high density of defects in both grain volume and on grain boundaries, which act as short circuit diffusion paths that may control and accelerate coalescence and grain growth kinetics, contributing to mechanical softening of the material. This point has to be verified experimentally with further investigations of the microstrural evolution of the alloys during the annealing treatment.

**Conclusion**

The aging behaviour of Fe-Si-Ti alloy processed by two kinds of SPD methods, cold-rolling and ECAP, was addressed in the present study. Severe plastic deformations usually offer good opportunity for enhancing the strain hardening potential in the fine grained structure and promote changes in the post-aging kinetics. In the present work, precipitation occurred quite quickly in the deformed material and precipitates appeared preferentially at grain boundaries. It was found that the strain hardening was mainly accommodated by localized precipitation at grain boundaries, which presumably contribute to reduce the effect of precipitation hardening in the as-deformed FeSiTi alloys. It was revealed that the two strengthening mechanisms combined in this study, by grain boundaries and precipitates lead to a lower increase in mechanical strength than the strengthening induced by the precipitation hardening acting alone in the non-deformed material.

**Acknowledgements**
This work is supported by the French National Agency (ANR) through functional materials and innovative processes program (project HYPERTUBE, n°ANR-09-MAPR-007).

**Tab. 1** Composition of the matrix and of the precipitates estimated from APT data for the FeSiTi alloy processed by rolling or ECAP and subsequently aged during 2 or 6 hours at 550°C. The volume fraction of precipitates was estimated using the nominal composition of the alloy as a reference

| Material processing | | Rolling state | | ECAP state | |
|---|---|---|---|---|---|
| Ageing times at 550°C | | 2h | 6h | 2h | 6h |
| Matrix (at.%) | Fe | 95.3 ±0.07 | 95.0 ±0.04 | 95.1 ±0.03 | 96.0 ±0.03 |
| | Si | 4.6 ±0.07 | 4.5 ±0.04 | 4.6 ±0.03 | 3.7 ±0.03 |
| | Ti | 0.1 ±0.01 | 0.3 ±0.01 | 0.2 ±0.01 | 0.2 ±0.01 |
| Precipitates (at.%) | Fe | 53.7 ±0.5 | 55.2 ±0.5 | 55.6 ±0.3 | 55.6 ±0.3 |
| | Si | 22.4 ±0.4 | 24.1 ±0.4 | 23.2 ±0.3 | 23.0 ±0.3 |
| | Ti | 23.6 ±0.4 | 20.4 ±0.4 | 20.9 ±0.3 | 21.0 ±0.3 |
| Estimated precipitate volume fraction | | 6.1 ±0.4 | 5.9 ±0.2 | 6.0 ±0.2 | 8.1 ±0.2 |

# Figures

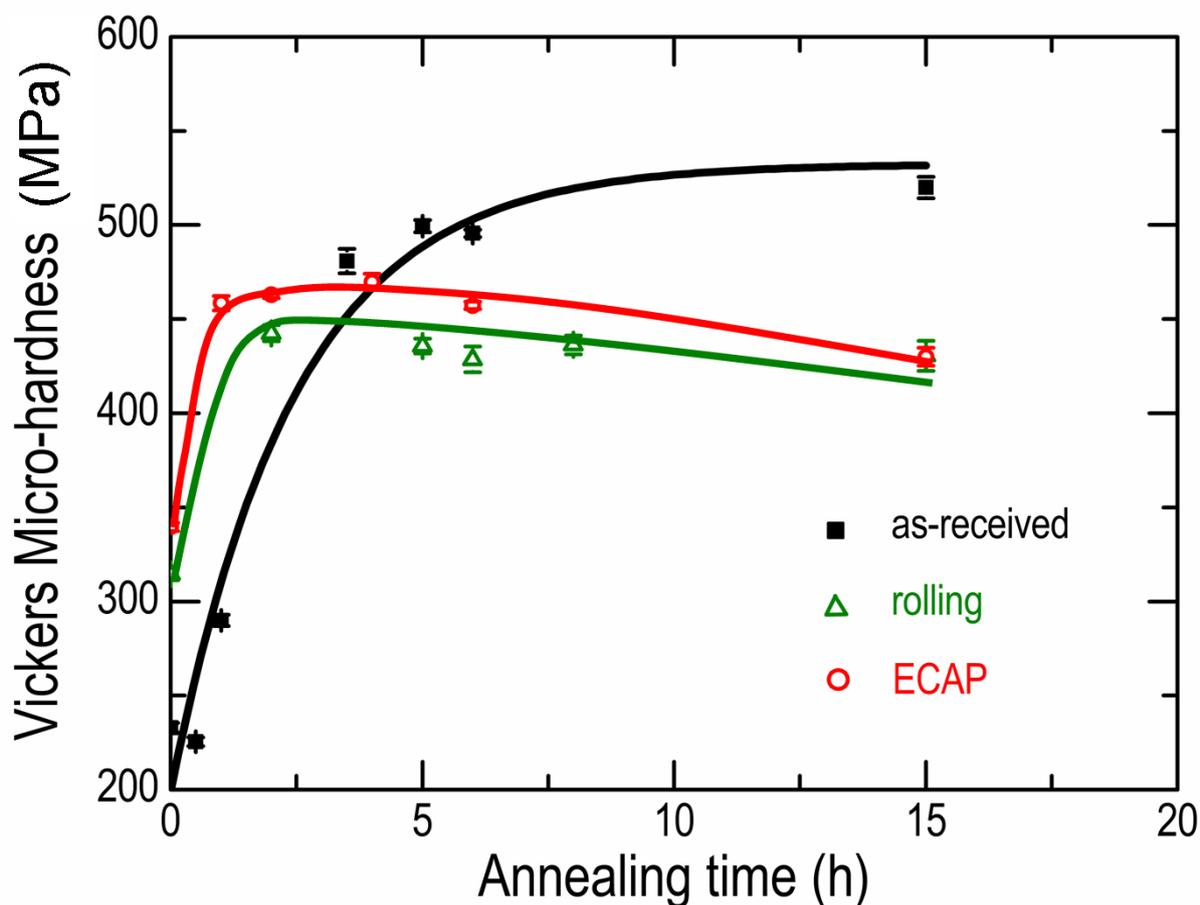

**Fig.1** Micro-hardness evolution of the FeSiTi alloy as a function of the aging time at 550°C for three different states: as-received, rolled, processed by ECAP

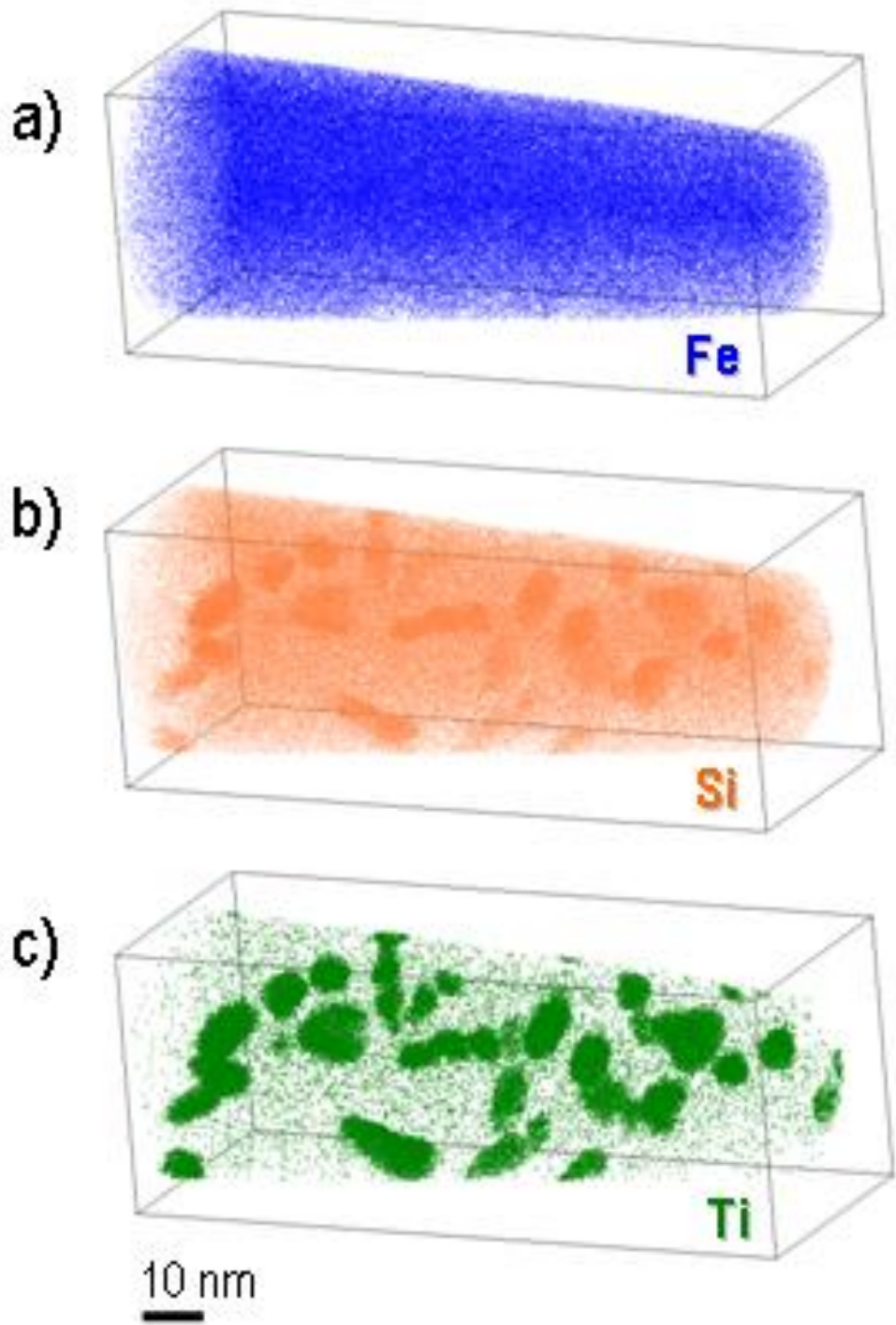

**Fig.2** FeSiTi alloy processed by ECAP and subsequently aged during 2h at 550°C. 3D reconstruction of a volume analysed by APT (43 x 43 x 111 nm$^3$). The distribution of the major elements is shown in three different images: Fe **(a)**, Si **(b)** and Ti **(c)**. Nanoscaled precipitates enriched in Ti and Si are clearly exhibited

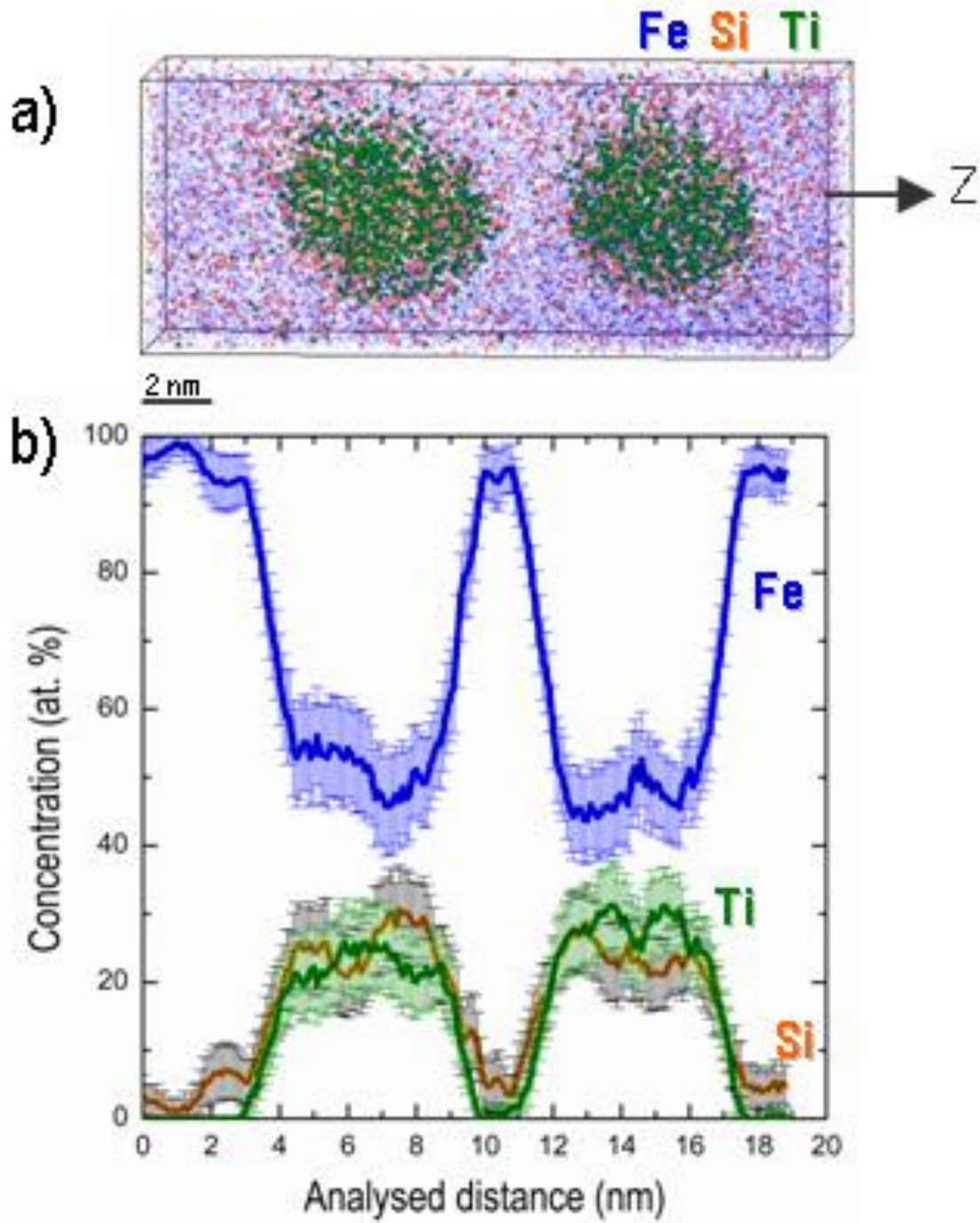

**Fig.3** FeSiTi alloy processed by ECAP and subsequently aged during 2h at 550°C. **(a)** Selected part (8 x 8 x 20 nm$^3$) of the 3D reconstruction of Fig. 2 showing two nanoscaled Si and Ti rich precipitates. **(b)** Concentration profile, derived from the reconstructed volume **(a)**, was drawn by moving a 1nm thick sampling slice by steps of 0.1nm along the Z direction, and show that precipitates do contain about 50%Fe, 25%Si and 25%Ti. Error bars are calculated from statistical sampling error inherent from the APT technique: standard deviation $2\sigma$ with $\sigma = (C(1-C)/N)^{1/2}$, N being the number of detected ions contained in the sampling box, and C the solute concentration

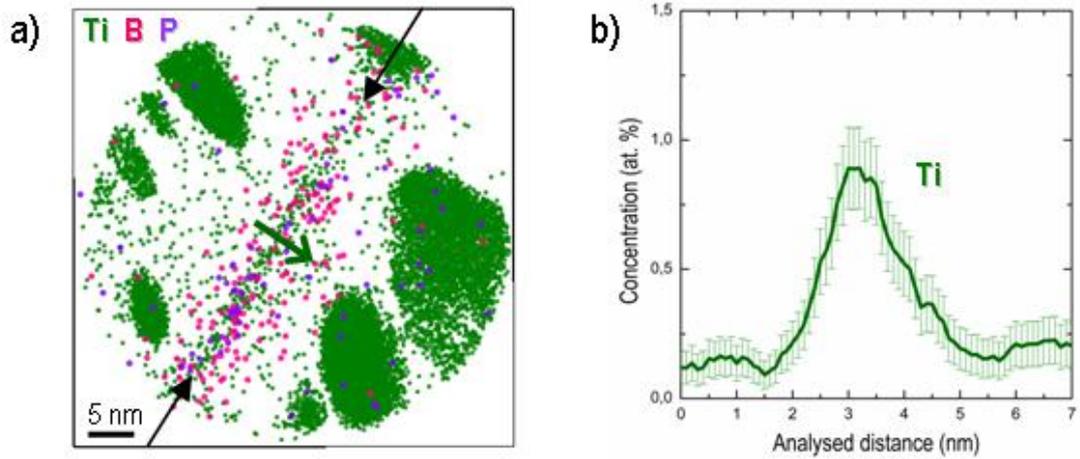

**Fig.4** FeSiTi alloy processed by rolling and subsequently aged during 2h at 550°C. **(a)** Cross section of the 3D reconstruction of a volume analysed by APT (48 x 48 x 13 nm$^3$). For more clarity, only Ti (green), B (pink) and P (purple) atoms are plotted. Ti rich nanoscaled precipitates are clearly exhibited. Beside, some Ti together with most of P and B are segregated along a narrow line that could be a grain boundary or a dislocation. **(b)** Ti concentration profile computed across the segregation line, drawn by moving a 0.8nm thick sampling slice by steps of 0.1nm along a direction perpendicular to the segregation (see green arrow). Error bars are calculated from statistical sampling error inherent from the APT technique: standard deviation 2σ with σ = (C(1-C)/N)$^{1/2}$, N being the number of detected ions contained in the sampling box, and C the solute concentration

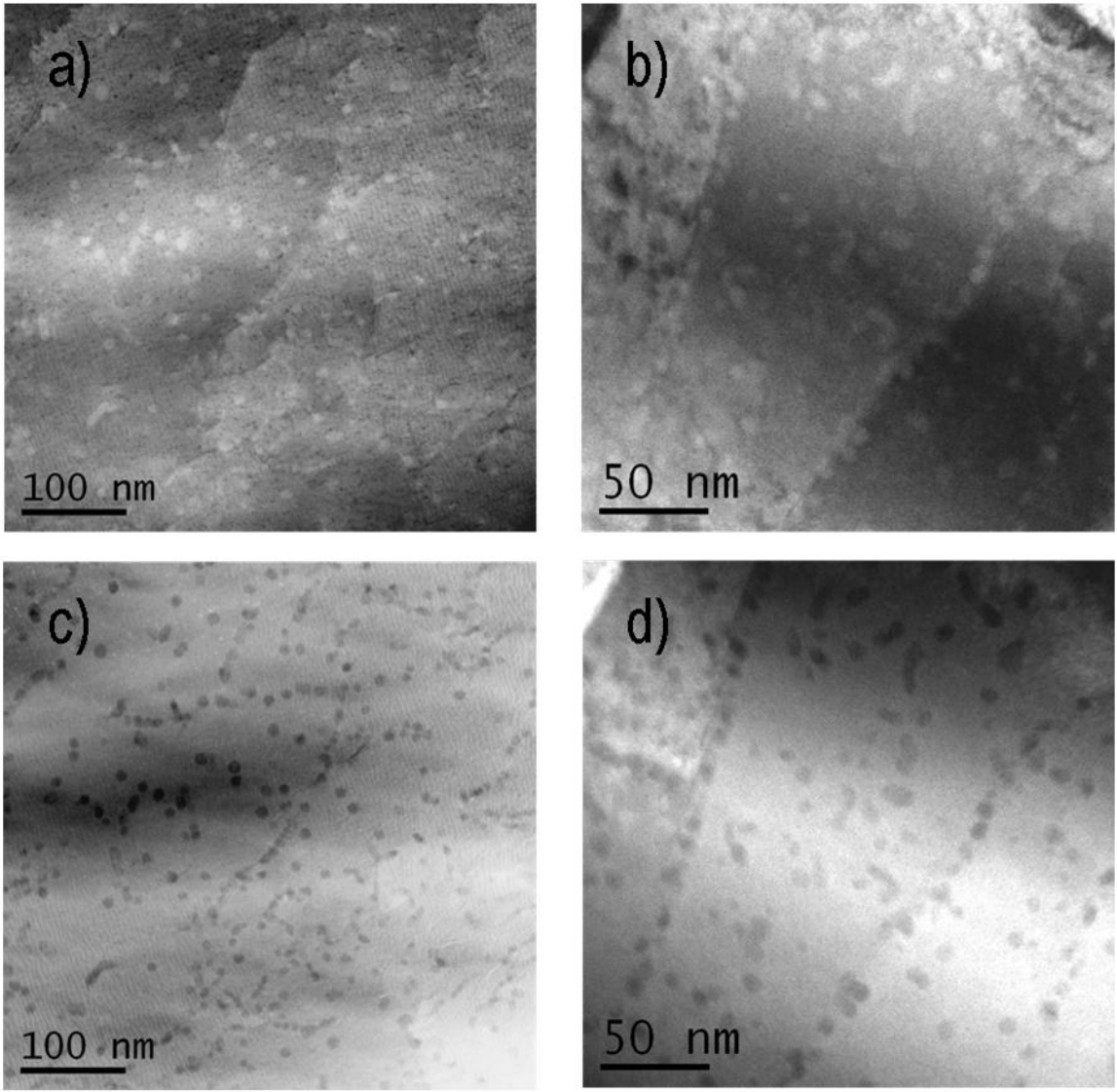

**Fig.5** BF-STEM (**(a)** and **(b)**) and corresponding HAADF-STEM (**(c)** and **(d)**) images of the FeSiTi alloy aged two hours at 550°C after rolling (**(a)** and **(c)**) and ECAP (**(b)** and **(d)**). Si and Ti rich nanoscaled precipitates are darkly imaged in HAADF because the average atomic number is lower. These images clearly show that the precipitate distribution is not homogeneous in both states. Some heterogeneous precipitation occurred along lattice dislocations or boundaries